\documentclass[twocolumn,showpacs]{revtex4}
\usepackage{graphicx}

\begin{document}

\title{Entanglement and nonlocality of a single relativistic particle}
\author{Jacob Dunningham$^{1}$ and Vlatko Vedral$^{1,2,3}$}
\affiliation{$^{1}$School of Physics and Astronomy, University of Leeds, Leeds LS2 9JT, United Kingdom \\ $^{2}$Centre for Quantum Technologies, National University of Singapore, 3 Science Drive 2, Singapore 117543\\
$^{3}$Department of Physics, National University of Singapore, 2 Science Drive 3, Singapore 117542}

\begin{abstract}
Recent work has argued that the concepts of entanglement and nonlocality must be taken seriously even in systems consisting of only a single particle. These treatments, however, are nonrelativistic and, if single particle entanglement is fundamental, it should also persist in a relativistic description. Here we consider a spin-1/2 particle in a superposition of two different velocities as viewed by an observer in a different relativistically-boosted inertial frame. We show that the entanglement survives right up to the speed of light and that the boosted observer would see single-particle violations of Bell's inequality. We also discuss how quantum gates could be implemented in this way and the possible implications for quantum information processing.
\end{abstract}

\maketitle

Entanglement and its related nonlocality are believed to be the root cause of all the major differences between quantum and classical physics. At present, however, nature requires two different theories to be combined in order to reach a satisfactory model of reality. Relativity is as important and well tested in its own domain as quantum mechanics and only the marriage of the two - known as quantum field theory - yields experimentally satisfactory results. It is therefore paramount that entanglement is analyzed from the relativistic perspective. Here we show that the nonrelativistic concept of single particle entanglement \cite{Tan91,Hardy94} survives in quantum field theory. Furthermore, boosted observers also see a single particle violation of Bell's inequalities. We prove, however, that the amount of entanglement is dependent on the inertial frame. Though the relativistic correction to entanglement is small at small speeds, this effect may play an important role in the future of quantum information processing.

Imagine that we have a massive particle of spin $s$ moving with a certain velocity $v_1$. When this particle is viewed by a relativistic observer traveling at velocity $v_2$ in the direction perpendicular to both $s$ and $v_1$, the effect is that the spin is rotated by an amount depending on the values of $v_1$ and $v_2$. The unitary matrix representing this rotation was worked out by Wigner in a seminal paper in $1939$ \cite{wigner}.
This spin rotation can be understood from the mathematical fact that a combination of two consecutive Lorentz boosts is  not itself a Lorentz boost. The extra transformation that is required is the Wigner rotation. More specifically if we perform a boost in the $x$ direction followed by a boost in the $y$ direction, the Wigner rotation will result in the $x-y$ plane.

Intuitively, the Wigner rotation can be explained as follows (see Figure 1). Imagine that a massive particle moves in the $y$ direction with velocity $v_1$ and its spin pointing in the $z$ direction (in the particle's reference frame). Introduce now an observer moving with velocity $v_2$ in the $x$ direction, i.e. perpendicular to both the spin and momentum of the particle. For him to calculate the value of the spin of the particle he needs to boost in the $x$ direction first and then apply boost in the $y$ direction. The resulting motion is not linear and gives rise to an effective angular momentum, which then couples to the spin (in much the same way that atomic spectra are affected by the spin-orbit coupling in elementary quantum mechanics). It is this coupling (that clearly depends on both $v_1$ and $v_2$) that leads to Wigner's rotation.

\begin{figure}[b] 
\includegraphics[width=6.5cm]{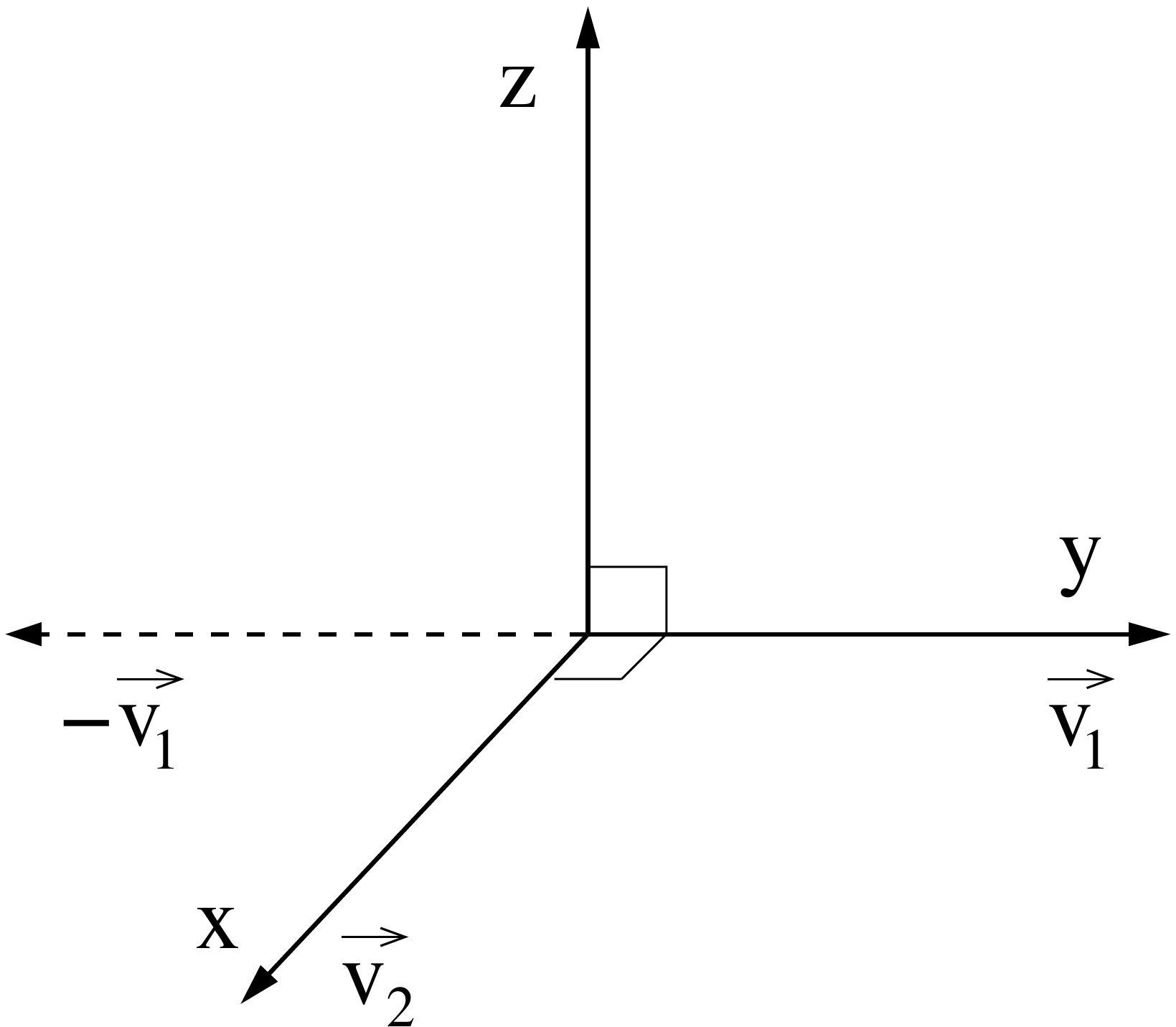}
\caption{A single spin-1/2 particle initially has its spin pointing up in the $z$-direction. It is boosted to some velocity $v_1$ in the $y$-direction (or a superposition of velocities $+v_1$ and $-v_1$ along the $y$-axis). The observer of the system is moving at velocity $v_2$ along the $x$-axis.} \label{axesfig}
\end{figure}

Unitary representations of Lorentz boosts and Wigner's rotations are part of the common folklore of quantum field theory and we need not explain them in detail here: an excellent treatment can be found in  \cite{Weinberg}. We immediately specialise to single particle states and confine all the formalism to this case only. For this we need to know that a general state of a single particle,
\begin{equation}
|\Psi \rangle = \int d \mu (v) f(v) |v\rangle |\chi\rangle
\end{equation}
where $d\mu$ is a relativistically invariant integration measure, $|v\rangle$ is the ket representing velocity, $|\chi\rangle$ is the ket representing spin, and $f(v)$ is the velocity space wavefunction, will transform under a general Lorentz boost in the following way:
\begin{equation}
U(\Lambda) |\Psi \rangle = \int d \mu (v) f(\Lambda^{-1}v) |v\rangle D(W(\Lambda))|\chi\rangle
\end{equation}
where $\Lambda$ is the Lorentz boost and $D(W(\Lambda))$ is the unitary transformation representing the Wigner rotation $W$ that itself is a function of the boost. We need not give the exact form of $D$ here as one specific example will be presented shortly.

In previous work \cite{Dunningham} we argued that the notion of single particle nonlocality (and, therefore, entanglement) should be taken seriously. To put it simply, a single particle existing in a superposition of two spatially distinct locations can violate Bell's inequalities in much the same way that the usual two particle EPR (or Bohm) state does. The main subtlety in this argument was that certain operations needed to be performed which appeared to contradict superselection rules. We do not wish to restate our arguments here, but, in short, a careful choice of ancillary systems allows us to sidestep any superselection restrictions.  The reader interested in a more detailed argument is encouraged to consult our discussion in \cite{Dunningham}.

We would now like to directly investigate the effects of Lorentz boosts on single particle entanglement and nonlocality. To put it more physically, if one observer records in his experiments a Bell violation due to single particle entanglement, will this also be true for all other inertial observers? Or, can one observer see something as entangled that appears disentangled to another observer who moves uniformly with respect to him?

While standard two particle entanglement has been studied a number of times with respect to relativity in inertial \cite{Peres2004, Gingrich2002, Li2003, Pachos2003a} and accelerated \cite{Fuentes2005a} frames, to our knowledge no-one has ever considered the case of a single particle. The importance of the latter are twofold. First, the relativistic behaviour is most transparent for single particles and any many-particle treatment follows straightforwardly by direct iteration of the single particle formalism. Secondly, it would be hard to argue
that single particle entanglement is genuine if it was only present in standard non-relativistic quantum mechanics.

The overall importance of studying entanglement in quantum field theory is itself beyond doubt. The ultimate information processors (as far as the contemporary opinion is concerned) are based on our most accurate description of bits and must  therefore be rooted in quantum field theory. Whether this gives us more power than the non-relativistic quantum computer remains to be investigated in more  detail.

We begin by taking the simplest single particle entangled state, that of the particle moving in two opposing directions along the $y$-axis with equal amplitudes and speeds, $v_1$. We consider a spin-1/2 particle that has two possible $z$-components of spin: spin-up, $|\uparrow\rangle$, and spin-down, $|\downarrow\rangle$. The spin of the particle is taken to initially point up in the $z$ direction and is  decoupled from momentum. This state can be written as
\begin{equation}
|\Psi\rangle  = \frac{1}{\sqrt{2}}(|v_1\rangle + |-v_1\rangle)|\uparrow\rangle. \label{singlepart}
\end{equation}
An observer boosted in the $x$ direction will see this state as
\begin{eqnarray}
|\Psi'\rangle  & = & \frac{1}{\sqrt{2}}| v_1\rangle (\cos \omega |\uparrow\rangle +i\sin\omega |\downarrow\rangle) \nonumber \\& + & \frac{1}{\sqrt{2}}|-v_1\rangle (\cos \omega |\uparrow\rangle - i\sin\omega |\downarrow\rangle) \label{psiprime}
\end{eqnarray}
where $\omega$ is the angle of Wigner's rotation given by,
\begin{eqnarray}
\sin\omega = \sqrt{\frac{(\gamma_1 -1)(\gamma_2-1)}{2(1+\gamma_1\gamma_2)}},
\end{eqnarray} 
with $\gamma_{1,2} = (1-(v_{1,2}/c)^2))^{-1/2}$. It should be immediately clear that the overall entanglement in this state (spin plus velocity) remains the same for the simple reason that the Lorentz boost is a local unitary transformation and therefore preserves overall entanglement. This, however, is no longer true for entanglement in the velocity degrees of freedom only.

\begin{figure}[t] 
\includegraphics[width=9cm]{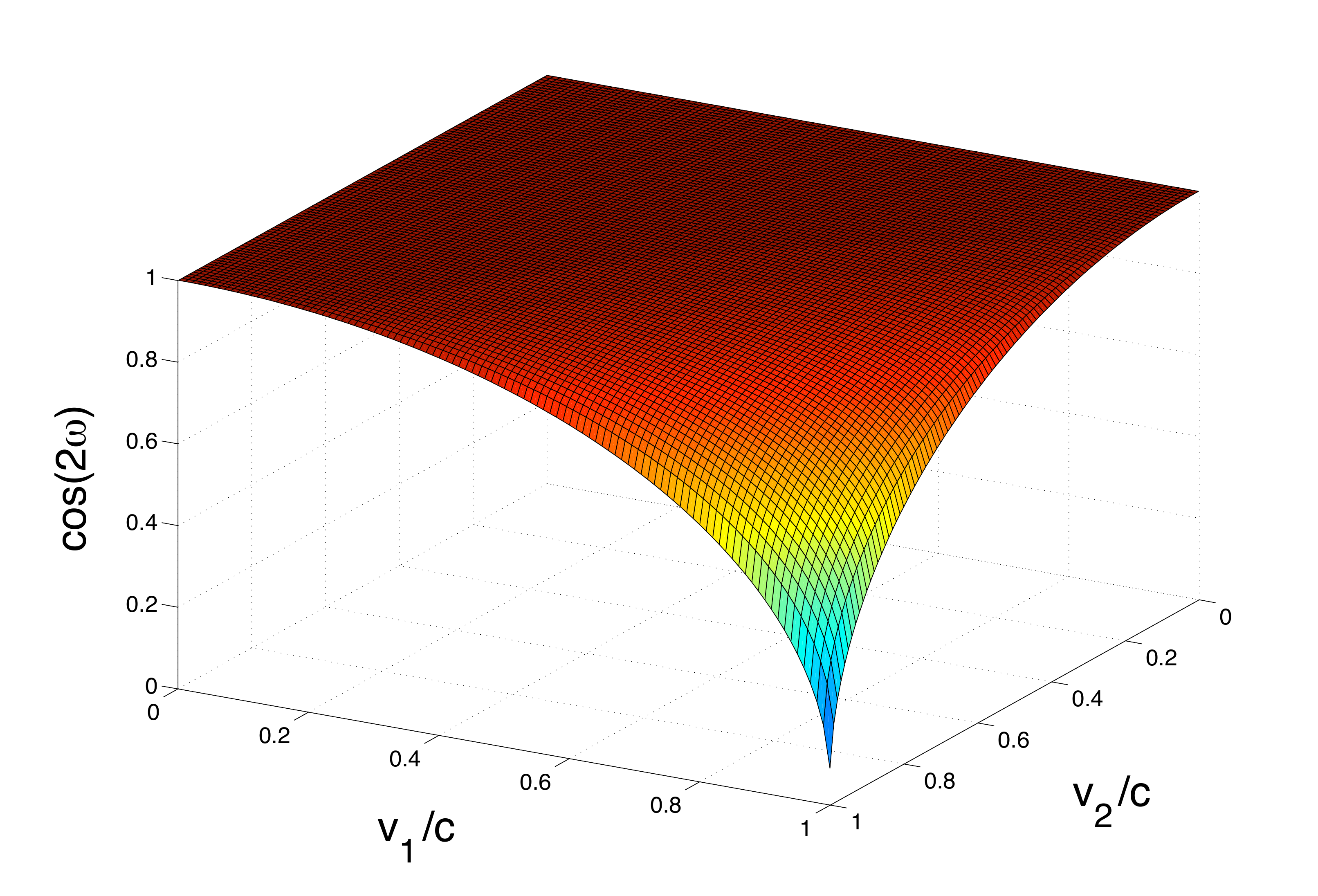}
\caption{The off-diagonal elements, $\cos(2\omega)$, of the reduced density matrix $\rho'$ as a function of the velocities $v_1$ and $v_2$.} \label{offdiag}
\end{figure}

In order to calculate entanglement in the velocity degrees of freedom of the particle we need to trace out the spin which, for the boosted observer, is now itself entangled to velocity. After this has been performed, the resulting density matrix is
\begin{eqnarray}
\rho'  &=&  \frac{1}{2} (|v_1\rangle\langle v_1|  + |-v_1\rangle\langle -v_1|   \nonumber \\ 
&+&  \cos (2\omega) |v_1\rangle\langle -v_1|  +\cos (2\omega) |-v_1\rangle\langle v_1|), \label{reddens}
\end{eqnarray}
where $\cos(2\omega)$ has the simple form
\begin{equation}
\cos(2\omega) = \frac{\gamma_1 + \gamma_2}{1+\gamma_1\gamma_2}.
\end{equation}

\begin{figure}[t] 
\includegraphics[width=9cm]{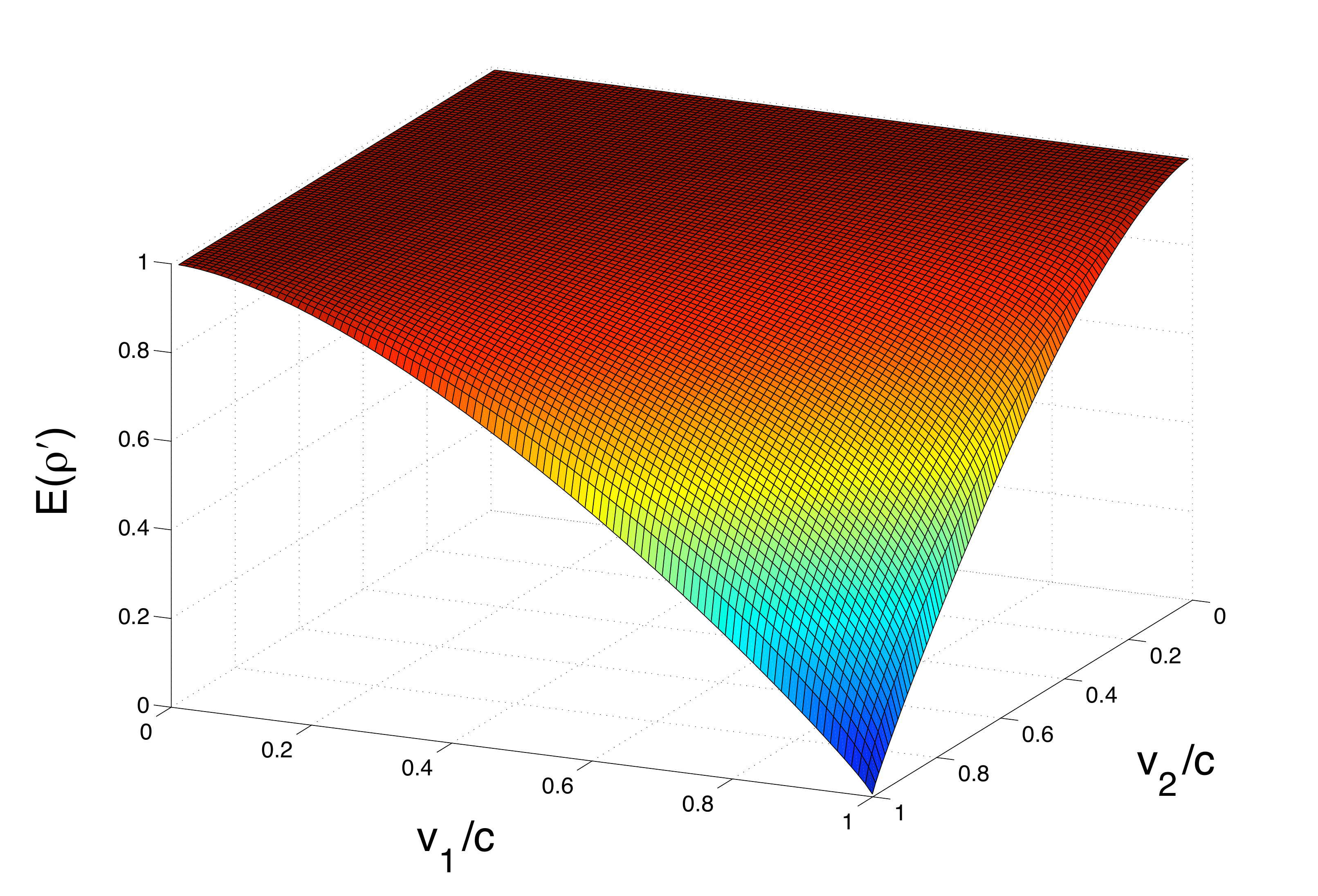}
\caption{The relative entropy of entanglement for $\rho'$ given by (\ref{reddens}) as a function of the velocities $v_1$ and $v_2$.} \label{entropy}
\end{figure}

From (\ref{reddens}) it is clear that the factor $\cos 2\omega$ determines the degree of the reduction of the off-diagonal elements, and hence the decoherence of entanglement. 
In the limit of small velocities we have $\gamma_1$, $\gamma_2 \to 1$, which means $\cos 2\omega \to 1$ and the state is maximally entangled. In the opposite limit, i.e. both velocities approach the speed of light, we have $\gamma_1$, $\gamma_2 \to \infty$, which means $\cos 2\omega \to 0$ and the state is disentangled. A plot of $\cos 2\omega$ as a function of $v_1$ and $v_2$ is shown in Figure~\ref{offdiag} and contains all the information we need to calculate the dependence of entanglement on relativity.

Firstly, we can calculate the relative entropy of entanglement for this state exactly. A simple calculation yields:
\begin{equation}
E(\rho') = 1 - S(\rho')
\end{equation}
where $S(\rho') = -{\rm tr}\left\{ \rho'\log_2 \rho'\right\}$ is the von Neumann entropy of the state $\rho'$ given by (\ref{reddens}).  This can be written as,
\begin{eqnarray}
E(\rho') = 1&+& \left(\frac{1+\cos 2\omega}{2}\right)\log_2 \left(\frac{1+\cos 2\omega}{2}\right)\nonumber  \\
&+&  \left(\frac{1-\cos 2\omega}{2}\right)\log_2 \left(\frac{1-\cos 2\omega}{2}\right).
\end{eqnarray}
A plot of $E(\rho')$ versus $v_1$ and $v_2$ is shown in Figure~\ref{entropy}. This has the same characteristics as the plot of $\cos 2\omega$. In particular, for small velocities, the state is maximally entangled and as the velocities approach the speed of light, the state becomes disentangled. Interestingly, the entanglement only vanishes when $v_1$ and $v_2$ are both equal to the speed of light. In other words, for massive particles the state will always appear entangled regardless of the boost applied to the observer.

Secondly, we can calculate the degree of violation of the CHSH version of Bell's inequalities. This is important since it is a common route to measuring the degree of entanglement in one or two particle systems. Here we rely on the result of the Horodecki family given in \cite{Horodecki}. The value of the Bell operator turns out to be
\begin{equation}
B = 2\sqrt{1+ \cos^2 2\omega}. \label{bellop}
\end{equation}
Violations of the CHSH inequality are possible for states for which $B\geq 2$.
The form of (\ref{bellop}) means that violations should be observable for our single particle state (\ref{singlepart}) for all observers right up to the speed of light.

Our treatment has only been for particles with mass. However, similar results can also be obtained for massless particles \cite{Alsing2002, Gingrich2004a}. In this case, the polarization of the photon takes the place of spin. Suppose, for example we had a single photon of a certain momentum that was polarized in the horizontal direction. This polarization can be written as the symmetric combination of different helicities, i.e. the left and right circularly polarized states. Under a general Lorentz boost, each term acquires a phase that depends on the helicity and the direction (but not the magnitude) of the momentum of the photon \cite{caban}. If the photon were put into a superposition of two different directions of momentum by, for example, a beam splitter, each of these momenta would be coupled to a different polarization state under a Lorentz boost. This leads to entanglement between the momentum and polarization states.

Our analysis firmly shows that the single particle entanglement is a genuine feature of quantum field theory and survives the introduction of relativity. It would be very intriguing to investigate how our results generalise to many particle relativistic quantum systems. What kind of many-body entanglement can be generated by Lorentz transformations? And how do these change in accelerated frames \cite{Fuentes2005a}?  More specifically, while the observer in the reference frame of the particle sees no entanglement between spin and velocity in the state $|\Psi \rangle$, the relativistic observer does see a certain amount of entanglement generated depending on the strength of the boost. The boost in this sense is equivalent to a controlled operation between spin and velocity qubits. 
For example, as $v_1$ and $v_2$ approach the speed of light, $\sin(\omega) = \cos(\omega) = 1/\sqrt2$, and the initial state (\ref{singlepart}) is seen by the observer as,
\begin{eqnarray}
|\Psi'\rangle  & = & \frac{1}{{2}}| v_1\rangle ( |\uparrow\rangle +i |\downarrow\rangle)  +  \frac{1}{{2}}|-v_1\rangle ( |\uparrow\rangle - i |\downarrow\rangle).
\end{eqnarray}
This can be viewed as a CNOT operation since, by boosting to the speed of light, the spin state $|\uparrow\rangle$ is transformed to one of two orthogonal states controlled by the velocity state, $|v_1\rangle$ or $|-v_1\rangle$.

It is common knowledge that once we can do certain controlled gates, any other computation is also possible. Can it, therefore, be that by boosting we can perform a wide range of quantum computations?  Bizarre though this may seem, it could well be that the computers of the future will utilise relativistic entanglement to perform gates in much the same way as presented here.

{\bf Acknowledgements:}
This work was financially supported by the United Kingdom EPSRC and the Royal Society and Wolfson Foundation
in the United Kingdom as well as the National Research Foundation and Ministry of Education, in Singapore.

\end{document}